\documentclass[prd, twocolumn, nofootinbib]{revtex4}

\usepackage{epsfig} 

\newcommand{\beq}{\begin{equation}}
\newcommand{\eeq}{\end{equation}}
\newcommand{\beqa}{\begin{eqnarray}}
\newcommand{\eeqa}{\end{eqnarray}}
\newcommand{\om}{\Omega_m}

\newcommand{\dlw}{\Delta w}

\def\fun#1#2{\lower3.6pt\vbox{\baselineskip0pt\lineskip.9pt
  \ialign{$\mathsurround=0pt#1\hfil##\hfil$\crcr#2\crcr\sim\crcr}}}

\begin{document} 

\title{How many dark energy parameters?} 
\author {Eric V. Linder$^1$ and Dragan Huterer$^2$} 
\affiliation{$^1$Physics Division, Lawrence Berkeley Laboratory, Berkeley, CA 94720\\
$^2$Kavli Institute for Cosmological Physics, and 
Astronomy and Astrophysics Department, University of Chicago, 
Chicago, IL 60637}

\date{\today} 

\begin{abstract} 
For exploring the physics behind the accelerating 
universe a crucial question is how much we can learn about the 
dynamics through next generation 
cosmological experiments.  For example, in defining the dark energy 
behavior through an effective equation of state, how many parameters 
can we realistically expect to tightly constrain?  Through both 
general and specific examples (including new parametrizations and 
principal component analysis) 
we argue that the answer is 42 -- no, 
wait, two.  Cosmological parameter analyses involving a 
measure of the equation of state value at some epoch (e.g.\ $w_0$) 
and a measure of the change in equation of state (e.g.\ $w'$) 
are therefore realistic in projecting dark energy parameter constraints. 
More elaborate parametrizations could have some uses (e.g.\ testing for 
bias or comparison with model features), but do not lead 
to accurately measured dark energy parameters.
\end{abstract} 


\maketitle

\section{Introduction} \label{sec.intro}

The discovery of the acceleration of the cosmic expansion has thrown 
physics and astronomy research into a ferment of activity, from a 
search for fundamental theories to investigation of parametrizations 
relating models and the cosmological dynamics to development of 
astrophysical surveys yielding improved measurements.  The question 
of the nature of the accelerating mechanism (``dark energy'') 
impacts the composition of some 70\% of the energy density of the universe, 
the evolution of large scale structure, and the fate of the universe. 
Learning the physics responsible will advance the fundamental 
frontiers of science, in high energy physics, extra dimensions, 
gravity beyond Einstein relativity, or possibly the unification 
of gravity and quantum physics. 

Given the open landscape of proposed theories, it is difficult to 
assess the prospects for a definitive measurement of 
the physics.  We consider here how much information the next generation 
of cosmological measurements is likely to realistically provide on the 
nature of dark energy.  By ``realistically'', we mean several things. 
Knocking down theories one by one is unlikely to be useful, given 
theorists' fecundity and fickleness.  Direct reconstruction, whether of the 
dark energy equation of state, density, or potential, while formally 
possible (in a limited range at least), is stymied by statistical and 
systematic errors in the observations.  So the question becomes, 
while describing dark energy in as model independent fashion as possible, 
what is the greatest degree of informative parametrization justified by 
the future observations. 

In \S\ref{sec.multipar} we examine various parametrizations, including 
designing some new ones useful for studying aspects of the dark energy 
dynamics, utilizing one to four parameters.  In \S\ref{sec.success} we 
define success criteria for a parameter estimation and project what 
constraints future supernova distance, CMB, and weak lensing shear 
measurements can impose, analyzing the resulting maximum number of 
parameters tightly fit.  Discussion of trade offs between success and 
model independence, such as in the minimum variance approach, is another 
issue addressed.  We consider in \S\ref{sec.eigen} uncorrelated and 
non-parametric characterizations.  \S\ref{sec.concl} 
examines what the data requirements would be to obtain more 
information on the nature of dark energy and summarizes the conclusions 
on the viable number of dark energy parameters.

\section{Modeling Dark Energy} \label{sec.multipar} 

Dark energy appears in the Friedmann equations of cosmological dynamics 
through its effective energy density and pressure.  We emphasize that 
these need only be effective quantities and not necessarily correspond 
to characteristics of a physical scalar field, say.  The ratio of the 
energy density to pressure, known as the equation of state ratio $w$, 
has importance within the equations regardless of its physical origin. 
This has been emphasized by \cite{linjen} (see also \cite{sahni}), in 
that any deviation of the Hubble parameter $H=\dot a/a$ from the matter 
dominated behavior can be written in terms of an effective equation 
of state. 

So a function $w(a)$, where $a$ is the scale factor of the universe, 
is key to understanding the dynamics of the cosmological expansion, 
in particular its acceleration.  With only imperfect knowledge of the 
expansion history $a(t)$, through noisy distance measures over a finite 
redshift range, one cannot derive the full function $w(a)$.  The minimum 
characteristics one would seek to provide insight on the origin of the 
acceleration and nature of the dark energy would be a measure of $w$ 
at some epoch, e.g.\ today, and a measure of its dynamics, i.e.\ its 
change over time (see \cite{caldlin} for a demonstration of the 
strength of this approach; see our \S\ref{sec.concl} for a discussion 
of the role of sound speed).  In particular these could yield consistency or 
conflict with the two key properties of the cosmological constant as 
the explanation: $w$ unchanging and equal to $-1$. 

Within a realistic scenario of next generation observations, how much 
can we expect to learn about the dark energy function $w(a)$?  We 
consider three approaches to answering this question.  After reviewing 
the basis for the current two parameter model, in \S\ref{sec.4par} 
we attempt to generalize it with a phenomenological approach,
calling out a set of four characteristics of an evolving equation of state,
and then examining restrictions of the phase space.  In \S\ref{sec.3par} 
we instead extend the two parameter model with physically motivated third 
parameters.  Later, for a more dark energy model independent tack, in 
\S\ref{sec.eigen} we investigate a principal component approach.

\subsection{Describing Dark Energy Evolution} \label{sec.4par} 

The two basic characteristics of a dynamical function mentioned above -- 
a value and a change in value -- can be implemented in many ways.  The 
one in perhaps most common use presently is employing the present value 
$w_0$ and the logarithmic derivative with respect to scale factor, $w'\equiv 
dw/d\ln a$, at redshift $z=1$ ($a=1/2$): 
\beqa 
w(a)&=&w_0-2w'(1-a) \nonumber \\ 
&=&w_0+w_a(1-a). \qquad {\rm [Model\ 2.0]}
\eeqa 
This possesses many useful properties, including a close approximation 
to a wide variety of dark energy dynamics, boundedness at early times 
(so CMB observations can be readily treated), analyticity of the Hubble 
parameter expression (involving an integral over $w$), and ease of 
physical interpretation \cite{linprl,lingrav}.  

To go beyond this description of 
dark energy dynamics, one might characterize the physics in a 
broadly phenomenological manner as having some equation of state value 
far in the past, $w_p$, (deep in the matter dominated epoch, near the time of 
CMB last scattering, say), some value far in the future, $w_f$, (deep in the 
dark energy dominated epoch), and treat the transition between the 
two as occurring at some scale factor $a_t$ and with some rapidity $\tau$. 

We choose to treat the transition in terms of the e-fold variable 
$N=\ln a$, since this is the characteristic scale of the cosmological 
background evolution.  That is, 
any dynamics driven purely by the expansion will have a transition scale 
length of order one in this variable.  If the equation of state changes 
according to $\dot w\sim H$, then $w'=dw/dN\sim 1$. 

In the neighborhood of the transition we can write 
\beq 
w(N)\approx w_t\left[1+(N-N_t)\frac{1}{w_t}\frac{dw}{dN_t} 
\right]. \label{eq.nearnt} 
\eeq 
There are many possible ways to extend this to a full function 
describing the past and future evolution of the equation of state. 
We would like this to be bounded in both directions (i.e.\ the past 
$N\ll N_t$ and the future $N\gg N_t$).  We choose to view Eq.\ 
(\ref{eq.nearnt}) as the first order term in an exponential, and 
applying our boundedness criteria we adopt 
\beq 
w(N)=w_f+\frac{\Delta w}{1+e^{(N-N_t)/\tau}}, \qquad {\rm 
[Model\ 4.0]} \label{eq.m40}
\eeq 
where $\Delta w=w_p-w_f$.  Note that $\tau=\dlw/[4(-dw/dN_t)]$. 

This is similar to the four parameter equation of state proposed by 
\cite{ccl}, 
\beq 
w(a)=w_0+(w_m-w_0)\frac{1+e^{a_c/\Delta}}{1-e^{1/\Delta}}\frac{1-e^{(1-a)
/\Delta}}{1+e^{(a_c-a)/\Delta}}, \label{eq.ccl} 
\eeq 
due to a similar 
phenomenology, but Model 4.0 has some advantages.  The previous version 
carried out the transition in $a$ rather than $N$ (also cf.\ \cite{bassett}), 
and so there was no 
natural scale for the transition rapidity, unlike the present case.  Also, 
Model 4.0 allows integration over $w$ to be done analytically and so 
the Hubble parameter can be written explicitly.  Note that 
Eq.\ (\ref{eq.m40}) is equivalent to 
\beq 
w(a)=w_f+\frac{\dlw}{1+(a/a_t)^{1/\tau}}, \qquad {\rm [Model\ 4.0]} 
\eeq 
and so 
\beqa 
H^2/H_0^2&=&\Omega_m a^{-3}+(1-\om)(1+a_t^{1/\tau})^{-3\tau'}\times 
\nonumber\\ 
&&\ a^{-3(1+w_f)} [1+(a/a_t)^{-1/\tau}]^{3\tau'}, 
\eeqa 
where $\om$ is the dimensionless matter density, $\tau'=\tau\dlw$, and 
we assume a spatially flat universe (as throughout this article). 

Our complete set of cosmological parameters for distance data then 
becomes $\{{\cal M},\om,w_f,\dlw,a_t,\tau\}$, where ${\cal M}$ is a 
nuisance parameter combining supernova absolute magnitude and the 
Hubble constant $H_0$.  The fourth of the dark energy parameters, 
measuring the rapidity, can be chosen to be $dw/dN_t$ rather $\tau$ if 
desired.  This is one new equation of state model we will consider; 
one unfortunate aspect is that Model 4.0 does not reduce to Model 2.0 
for any choice of parameters, so it cannot be viewed as an extension 
of the ``baseline'' Model 2.0.

\subsection{Extending Two Parameter Dynamics} \label{sec.3par} 

Given the successful properties of the two parameter model, and that 
extensive estimation of constraints on the two model parameters exists 
for present and next generation cosmological probes, let us 
attempt to extend Model 2.0 to a third parameter, while keeping 
the virtues as intact as possible.  Studies of many dark energy models 
in the $w'-w$ phase plane reveal interesting properties of the dynamics 
\cite{caldlin} and show the success -- and limitations -- of Model 2.0. 

This model keeps the derivative 
$dw/da$ constant, or the logarithmic derivative $w'\equiv dw/dN\equiv 
dw/d\ln a\sim a$. 
One might conjecture that the logarithmic derivative, giving the change 
in dark energy equation of state per e-fold of cosmic expansion, is an 
important characteristic of the dynamics \cite{linmiq}; so let us 
venture a generalization by enlarging its scope of behavior.  (A different 
approach in terms of changing the transition epoch is considered by 
\cite{rapetti}.)  We consider 
$w'\sim a^b$: 
\beq 
w(a)=w_0+w_a(1-a^b).\qquad {\rm [Model\ 3.1]} 
\eeq 
Since Model 2.0 is a fairly successful approximation for many dark 
energy models we might expect that $b$ does not deviate greatly from unity. 
In terms of the phase space dynamics of \cite{caldlin}, their ``thawing'' 
models are a special case of Model 3.1, with $w_a=-(1+w_0)$ and their 
slope equal to our $b$. 

Of course for a model with $w_a=0$ (such as taking the cosmological 
constant as the fiducial model), observations will place no constraints on 
the third parameter $b$.  Therefore, 
this is not a general extension of the two parameter model.  So we 
also consider a different way 
of changing the dynamics $w'$.  We introduce some scale factor 
dependence to $dw/da$ by adding a term of higher order in $1-a$: 
\beq 
w(a)=w_0+w_a(1-a)+w_3(1-a)^3. \qquad {\rm [Model\ 3.2]} 
\eeq 
Using the cubic power rather than a quadratic preserves the trend 
of the logarithmic derivative $w'$ at the present, $a=1$.  For 
example if $w_a=0$ then with this cubic the sign of $w'$ does not 
flip around the present day, unlike with a quadratic.  We make no 
claims, however, that the cubic form is 
of great significance; it merely represents a possible three parameter 
form with reasonable properties, extending the standard two parameter 
model. 

It is instructive to also investigate a more extreme evolution of the 
equation of state, such as 
\beq 
w(a)=w_0+w_a(1-a)+w_e[(1+z)e^z-1], \quad {\rm [Model\ 3.3]} 
\eeq 
to see the influence of runaway behavior at high redshifts.  This 
will not be suitable for treating CMB or growth related data, and we 
only consider it briefly. 

Below we give the expressions for the modifications to the Hubble 
expansion $\delta H^2=H^2/H_0^2-\om a^{-3}$, 
divided by $(1-\om)$, for the models considered: 
\beqa 
{\rm [Model\ 3.1]:}&&\!\!\!\!\!\! a^{-3(1+w_0+w_a)}e^{-3w_a(1-a^b)/b} \\ 
{\rm [Model\ 3.2]:}&&\!\!\!\!\!\! a^{-3(1+w_0+w_a+w_3)}\times \\ 
&&\!\!\!\!\!\! e^{-3w_a-11w_3/2+3(w_a+3w_3)a-9w_3a^2/2+w_3a^3} \nonumber \\ 
{\rm [Model\ 3.3]:}&&\!\!\!\!\!\! a^{-3(1+w_0+w_a-w_e)}e^{-3w_a(1-a)+3w_e(e^z-1)} 
\eeqa

\section{Searching for Successful Constraints} \label{sec.success} 

One question to address before further investigation is what is our 
criteria for success in constraining dark energy parameters.  We can 
obviously fit a model with many parameters, each poorly, but what we 
mean by ``how many dark energy parameters?'' is how many can we 
determine to an accuracy that provides real physical insight.  What 
the criteria for success should be is not a solved problem, though 
recent progress has been made by \cite{caldlin,linbw}.  

We adopt the following unrigorous but reasonable seeming criteria: 
1) a parameter describing the equation of state at some epoch should be 
determined to a fractional error of $<10\%$, relative to its fiducial 
value, e.g.\ the future equation of state value should be found to 
within 0.1, and 2) a parameter describing the change in equation of state, 
or a derivative, should be estimated to within 0.2 (i.e.\ 20\% of the 
Hubble time scale). It is hard to see that much looser constraints 
would be of substantial use in unraveling the physics. 

We carry out a full Fisher analysis of the cosmological parameter 
uncertainties to test how many equation of state parameters can be 
reasonably constrained, in addition to the matter density $\om$, 
and other parameters relevant to the cosmological measurements. 
Three types of cosmological probes are considered: Type Ia supernova 
distances (SN), weak lensing shear (WL), and the 
cosmic microwave background (CMB) angular power spectrum.
We employ supernova distance data of the quality expected from the 
next generation experiment SNAP \cite{snap}, with 2000 supernovae 
from $z=0.1-1.7$ plus 300 local supernovae from $z=0.03-0.08$, with 
intrinsic magnitude dispersion 0.15 magnitudes (7\% in luminosity 
distance) plus an irreducible systematic of $0.02(1+z)/2.7$ mag per 
0.1 redshift bin.  This method carries with it the parameter ${\cal M}$ 
to be marginalized over.   

For the fiducial weak lensing survey we assume sky coverage of 1000
sq.\ deg.\ with 100 galaxies per arcmin$^2$, which roughly corresponds to
coverage and depth expected from SNAP \cite{wliii}. We also briefly 
consider a LSST-type
survey with 15000 sq.\ deg.\ and 30 galaxies per arcmin$^2$. Throughout we
consider lensing tomography with 10 redshift bins equally spaced between $z=0$
and $z=3$ and use the lensing power spectra on scales $50\le \ell \le
3000$. Intrinsic shape noise of each galaxy is fixed to
$\sigma_\gamma=0.22$. In addition to the dark energy parameters (those
describing the equation of state, plus $\Omega_m$) we vary the the amplitude of
mass fluctuations with the fiducial value $\sigma_8= 0.9$, the physical matter
and baryon densities $\Omega_m h^2=0.147$ and $\Omega_b h^2=0.023$, 
and the scalar spectral index $n=1.0$. 
The sum of the neutrino masses is held fixed at $m_{\nu}=0.1$eV. 
We compute the linear power spectrum using the fitting formulae of 
Hu \& Eisenstein \cite{Hu_transfer}.  We generalize
the formulae to $w\neq -1$ by appropriately modifying the growth function of
density perturbations. To complete the calculation of the full nonlinear power
spectrum we use the fitting formulae of Smith et al.\ \cite{smith}. 
We emphasize that the calculations here are overoptimistic since we do not
consider any systematics in the weak lensing survey.  
(On the other hand, it is conceivable that additional information, 
not captured by the two-point correlation function, can and will be 
extracted from weak lensing maps.) 
Understanding of weak lensing systematics is at an early stage and we 
hope to include a more realistic treatment in the future; for preliminary
requirements on WL systematics see, e.g., \cite{wlsys}. 

The third piece of information we (optionally) add is the full Fisher matrix
corresponding to the expected constraints from the Planck CMB experiment with
temperature and polarization information \cite{EisHuTeg}.  In the SN+CMB case
(i.e. when WL information is not considered), the Planck information is
captured by the determination of the CMB peak locations, or measurement of the
angular diameter distance to the last scattering surface with a fractional
precision of 0.7\%.  As shown in \cite{frieman}, this nicely complements SN
measurements, typically being more powerful than an independent prior on
$\Omega_M$ coming from large-scale structure surveys.

\subsection{Fitting the Four Parameter Model} 

To provide a strong chance for tight parameter constraints on all four 
parameters in Model 4.0, we take a sensitive fiducial cosmological model 
with $w_f=-1$, $\dlw=0.8$ ($w_p=-0.2$), $a_t=0.5$, and $\tau=0.2$.  Note 
that the last value corresponds to $dw/dN_t=-1$.  So this model has a 
strong transition, at an epoch accessible to the data, over a time scale 
also visible to the data.  

To get a first look at how difficult the task is, we look at 
the unmarginalized uncertainties, i.e.\ the parameter estimations if 
all other parameters (for both the equation of state and $\om$, ${\cal M}$) 
are fixed.  In this case, SN+CMB data provide 
$\sigma(\om,w_f,\dlw,a_t,\tau)=(0.0043, 0.010, 
0.043,0.014,0.018)$, or 0.09 for $\sigma(dw/dN_t)$.  These all satisfy 
the success criteria (note the constraint on $\tau$ or $dw/dN_t$ would 
be borderline if we had applied blindly the 10\% fractional precision rule; 
for consistency, if we apply the 0.2 rule to $\dlw$ and 
$dw/dN_t$ then the precision requirement on $\tau$ should be $\sim20\%$). 

Other than 
the rapidity, the equation of state parameter estimations succeed by 
factors of 4-10.  Since this is not a huge margin, we have our first 
indication that it may be difficult to characterize accurately the 
equation of state with several parameters, given the loss in precision 
that marginalization would cause.  However we should also add in the 
WL data (though this is already marginalized over the large scale 
structure parameters) and now the (mostly) unmarginalized estimations 
are $\sigma(\om,w_f,\dlw,a_t,\tau)=(0.0030,0.011,0.065, 0.015, 0.016)$ 
for SN+WL and $(0.0029,0.0099,0.038, 0.13, 0.015)$ for SN+WL+CMB. 

We now examine this model in detail, 
gradually restricting the four parameter model to fewer parameters 
by fixing the others, one at a time, in a physically motivated manner. 
Full results appear in Table~\ref{table.4par}.  For the full four 
parameter model even all three probes together only allow one parameter 
($a_t$) to be satisfactorily fit.  This is despite the clear 
complementarity: for example on $a_t$, SN+CMB gives an uncertainty of 
0.25, SN+WL gives 0.17, SN+WL+CMB provides 0.045.   
We also note that moving away from the fiducial model 
can cause drastic increase in the uncertainties; for example with $\tau=1$ 
or $w_p=-0.8$ the parameter estimation errors exceed unity. 

So even with next generation measurements of distances to better than 
1\%, wide area weak lensing shear information, and accurate determination 
of the distance to the CMB last scattering surface we will be unable 
to accurately characterize the dark energy equation of state with 
four parameters.  In the next subsection we explore 
the situation for reduced parameter sets. 

\begin{center}
\begin{table}[!h] 
\begin{tabular}{l|c|c|c|c}
Model&$\sigma(w_f)$&$\sigma(\dlw)$&$\sigma(a_t)$&$\sigma(\tau)$\\ 
\hline 
{\it Success criterion}&{\it 0.1}&{\it 0.2}&{\it 0.05}&{\it 0.04}\\ 
\hline\hline 
All 4 parameters&0.15&0.30&0.045&0.13\\ 
\hline 
3 par: fix $w_f$&--&0.095&0.045&0.050\\ 
\hline 
3 par: fix $\dlw$&0.047&--&0.043&0.056\\ 
\hline 
3 par: fix $a_t$&0.15&0.29&--&0.13\\ 
\hline 
3 par: fix $\tau$&0.057&0.13&0.044&--\\ 
\hline 
2 par: fix $a_t,\tau$&0.048&0.13&--&--\\ 
\hline 
2 par: fix $\dlw ,\tau$&0.041&--&0.043&--\\ 
\hline 
2 par: fix $\dlw ,a_t$&0.027&--&--&0.056\\ 
\hline 
2 par: fix $w_f,\tau$&--&0.092&0.037&--\\ 
\hline 
2 par: fix $w_f,a_t$&--&0.053&--&0.041\\ 
\hline 
2 par: fix $w_f,\dlw$&--&--&0.025&0.049\\ 
\hline 
\end{tabular} 
\caption{Cosmological parameter sensitivities for the four parameter 
equation of state model are here estimated for the combination of 
next generation supernovae, weak lensing, and CMB data.  Fixed parameters 
are denoted with --.  
The second row shows the criteria for successful
determination of each parameter (see text for details). The 
matter density $\om$ estimation is always better than 0.01.} 
\label{table.4par}
\end{table} 
\end{center}

\subsubsection{Reduction to three parameters} \label{sec.4to3} 

Fixing parameters corresponds to using theoretical prejudice to 
pre-determine characteristics of the equation of state.  Ideally, 
this is physically motivated prejudice, but sometimes the choice 
is not clear and one should scan the phase space to see how the 
fiducial choice of the fixed parameter affects the results. 

First we address the variable $w_f$ describing the asymptotic future 
value of the equation of state.  Since observational data only exist 
for the past, we have no direct constraint on the future.  If the 
transition from past to future values is not complete, then the data 
should not be sensitive to the exact choice of $w_f$.  We fix 
$w_f=-1$, corresponding to an asymptotic deSitter state, arguably 
better motivated than some other value.  For the resulting three 
free parameter equation of state the uncertainties using SN+WL+CMB 
meet the criteria for success for only two parameters ($\dlw$ and 
$a_t$).  Again, changing the fiducial values tends to increase 
the errors dramatically. 

Next we fix the transition epoch $a_t=0.5$.  This could be roughly justified 
by the expectation that dark energy should start to show increased 
dynamics as its energy density begins to overtake the matter density. 
Note that the standard Model 2.0 evaluates $w'$ at $a=0.5$ for the 
same reason, and proves successful in matching many scalar field and 
extended gravity model behaviors.  Now the three free parameter 
equation of state model fails to fit any parameter satisfactorily (recall 
that $a_t$ was the one successful fit in the full four parameter case, 
and now we have rendered that moot).  Moving the transition epoch 
further in the past allows better determination of $w_f$, e.g.\ taking 
$a_t=0.3$ yields $\sigma(w_f)=0.053$ rather than 0.15, but raises the 
uncertainty on $\dlw$ and $\tau$.   Conversely, making the transition 
more recent improves estimation of the rapidity $\tau$ (and of the high 
redshift value $w_p$) but worsens other parameter errors. 

Fixing the rapidity $\tau$ or $dw/dN_t$ has less strong motivation, 
but one could argue that $dw/dN_t=-1$, i.e.\ evolution on the Hubble 
time scale, is a reasonable choice.  For the fiducial parameters this 
corresponds to $\tau=0.2$.  
With fixed $\tau=0.2$, the three free parameter
equation of state allows successful fit of all three parameters. 
But as we say, it is less obvious that we can justify 
holding the rapidity fixed, and if the fiducial is changed to $\tau=1$, 
then none of the parameters can be reasonably estimated.  

Finally, taking fixed the degree of change of the equation of state 
from the far past to far future, $\dlw=w_p-w_f$, again less physically 
justified, allows successful estimation of two of the three remaining 
equation of state parameters (for the extreme case $\dlw=0.8$). 

One would need to carry out a comprehensive scan of phase 
space (e.g.\ along the lines of the Markov chain Monte Carlo \cite{coramc} 
applied to the four parameter model Eq.~\ref{eq.ccl} in terms of 
scale factor $a$) to be sure that the reduced three parameter equation 
of state model cannot satisfy the characterization criteria in any 
particular case.  However we have shown that we cannot generically 
expect satisfactory three parameter fits, and indeed cannot obtain 
them (except in the fixed rapidity case) in a fiducial model designed 
to be especially favorable to see the behavior of the equation of state 
(we have checked that $\tau=0.2$ gives near optimal sensitivity).  
Moreover, since we do not know which fiducial case represents the true 
universe, we cannot claim three equation of state 
parameters are justified unless such a model is successfully fit over 
a broad region of phase space.

\subsubsection{Reduction to two parameters} \label{sec.4to2} 

Of course we know that it is possible to achieve a successful two 
parameter characterization of the dark energy equation of state: that 
is precisely what the standard Model 2.0 does.  For the data simulation 
used here, Model 2.0 delivers $\sigma(w_0,w')=(0.066,0.11)$ for a 
fiducial model of the cosmological constant. 

Reducing Model 4.0 to a two parameter case by fixing two parameters 
at a time gives six permutations.  These are presented in 
Table~\ref{table.4par}.  Briefly, all cases where $\tau$ is one of 
the parameters fixed can successfully fit two parameters (note that 
we also can find the derived parameter $w_p$, the high redshift value 
of the equation of state to better than 0.1).  However when $\tau$ is 
one of the two free parameters, we fail to estimate it well and so 
only attain a one parameter fit in the equation of state. 

Note that with only one of either supernova or weak lensing data, 
the only one of the six permutations of the doubly reduced four 
parameter model that even marginally succeeds in fitting two 
parameters is where we fix $a_t$ and $\tau$ -- the major part of the 
dynamics we seek to find!  This further demonstrates that it is 
remarkably difficult to characterize detailed dynamics of dark energy, 
even using next generation data, and complementarity is essential.

\subsection{Results from Three Parameter Extensions} \label{sec.3parres} 

If the full four parameter dynamics generalization fails, perhaps 
a more limited extension of the standard two parameter (value and 
derivative) approach would work, as proposed in \S\ref{sec.3par}. 

\subsubsection{Power law extension} 

In Model 3.1, we allowed $dw/da$ 
to depend on $a$, as a power law.  Recall that no constraint can 
be placed on the third parameter $b$ when the fiducial model has 
$w_a=0$.  Also, a fiducial value near $b=1$ (i.e.\ looking 
like the standard Model 2.0) does not change the constraints on the 
other equation of state parameters very much.  Taking $b$ to differ 
from unity by 10\% changes the uncertainties in $w_0$ and $w_a$ by 
roughly 10\% and 7\%.  However, even when using a fiducial model with 
$w_a\ne0$ it is not possible to obtain a reasonable constraint on $b$. 
In the SUGRA model with $w_0=-0.82$, $w_a=0.58$, the third parameter 
can only be estimated with $\sigma(b)\approx0.8$ (without all three 
probes this becomes worse than 2). 
So this model does not allow characterization of dark energy behavior 
beyond two parameters.  

\subsubsection{Cubic extension} 

In the cubic Model 3.2, we do not have the problem with no determination 
of the third parameter for fiducial $w_a=0$.  However now covariance 
with the additional parameter significantly degrades estimation of the 
first two parameters, especially $w_a$, unless tightly controlled by 
inclusion of all three probes.  With all three sources of information, 
for the fiducial cosmological constant case the uncertainty 
$\sigma(w_3)=0.64$.  While all three parameters can be satisfactorily 
fit for the SUGRA case (0.074 uncertainty in $w_3$), we have no guarantee 
that the universe lives in such a favorable case, and so we cannot 
generically gain a third parameter to teach us about dark energy. 

\subsubsection{Misleading sensitivity} 

To attempt to constrain a third parameter tightly enough to be useful, 
we are led to models with high sensitivity to the equation of state, 
e.g.\ by rapid variation.  Indeed this is basically what allows Model~3.2 
to succeed as well as it does: the cubic behavior, though bounded, gives 
strong evolution.  To carry the sensitivity to an extreme, we consider 
the exponential Model 3.3.  Even using only the single probe of 
supernovae distances, we find that indeed we can obtain successful 
constraints of $\sim0.2$ on $w_e$.  By contrast, with supernovae only, 
the third parameters in Models~3.1 and 3.2 can only be determined to 
$\sim10$. 

This enhanced sensitivity is a bug, not a feature.  The equation of 
state model is pathological, blowing up for high redshifts.  Indeed 
we cannot straightforwardly employ CMB data or weak lensing data (because 
it depends on the growth function from high redshift).  The same 
situation arises, though less extremely, in the pioneering, but now 
obsolete, parametrization first considered in \cite{wprime}: 
\beq 
w(z)=w_0+w_1z.\qquad{\rm [Model\ 2.1]} 
\eeq 
Even cutting this off at some moderate redshift and stitching on a 
more well-behaved parametrization can cause misestimation of cosmological 
parameter determination. 

For example, spurious estimations can result from using Model~2.1 
for baryon oscillation surveys.  These data typically extend to 
$z=1.5-3.5$ and sensitivities for baryon oscillations alone can 
appear quite promising.  Parameter estimations in terms of 
$w_1$ are factors of 1.9-2.7 times better than those in terms of 
$w'\equiv w_a/2$; however, as stated above this is a bug not a feature. 
This hypersensitivity arises from the extreme evolution forced upon 
the equation of state (even when the fiducial model is taken to be 
$w_1=0$, the form still enters in the Fisher derivatives).  One can 
test this by considering only low redshift ($z<1$) data, where 
Model~2.1 is not too egregious; indeed then the sensitivities in 
terms of $w_1$ and $w'=w_a/2$ agree.  This is not to say that baryon 
oscillations do not provide a possible cosmological probe -- they 
do in complementarity with other methods.  Rather it cautions against 
applying a parametrization outside its realm of validity.  Since we 
are now in the fortunate situation of employing data sets extending 
beyond $z\approx1$, Model~2.1 has become unsuitable and should be retired.

\subsection{Success vs.\ Model Independence} \label{sec.modelindep} 

It appears that obtaining a three parameter characterization of the 
dark energy equation of state is nontrivial.  We simultaneously need 
high sensitivity of the dark energy to the third parameter, and small 
covariance with the other parameters, so as not to increase their 
uncertainties.  We use this idea in \S\ref{sec.eigen} to make one more 
attempt at describing the equation of state in more detail. 

First we note that the equation of state parameter uncertainties 
calculated need not be the 
minimum errors.  For a given form of $w(a)$ one can look for a change 
of variables that minimizes the error on some variable, or some other 
quantity (such as volume of the N-dimensional uncertainty ellipsoid).  
However the price to pay for this improvement is breakdown of model 
independence.  The new variables formed will no longer have the same 
physical meaning as one changes the experiment (e.g.\ maximum redshift), 
cosmological probe, or fiducial model. 

To emphasize this point more strongly, we consider the case of 
minimum variance, or the ``pivot redshift'' (see \cite{huttur} 
and also Appendix A of the 
omnivaluable \cite{EisHuTeg}).  For the standard Model 2.0, say, we can 
ask what variable characterizing the value of the equation of state shows 
the minimum uncertainty.  The uncertainty on $w(a)$ is 
\beq 
\sigma^2(w(a))=\sigma^2(w_0)+(1-a)^2\sigma^2(w_a)+2(1-a)COV[w_0,w_a], 
\eeq 
where $COV$ represents an entry in the covariance matrix. 
Minimizing this with respect to $a$, to find the smallest uncertainty 
on the value of the equation of state at any epoch, we find that at 
$a_{min}=1+COV[w_0,w_a]/\sigma^2(w_a)$ the minimum uncertainty is 
obtained, on the new variable $w_{min}=w_0+w_a(1-a_{min})$: 
\beq 
\sigma^2(w_{min})=\sigma^2(w_0)-COV^2[w_0,w_a]/\sigma^2(w_a). 
\eeq 
This is manifestly smaller than $\sigma(w_0)$. 

But $w_{min}$ has no absolute meaning independent of probe, survey 
characteristics, priors, or fiducial model, and so it cannot be used to 
compare experiments or models.  For example, for the supernovae 
plus CMB data set, $z_{min}=0.28$ and $\sigma(w_{min})=0.038$.  But that 
was for a fiducial cosmological constant; if a fiducial SUGRA model is 
used then $w_{min}$ has a different meaning, and $z_{min}=0.52$ and 
$\sigma(w_{min})=0.016$.  If no CMB data is used, instead a 0.03 prior 
on the matter density, then the cosmological constant case has 
$z_{min}=0.10$ and $\sigma(w_{min})=0.072$. 

Since the main point of a phenomenological equation of state rather 
than an exact solution for a specific scalar field dynamics is the 
utility of model independence (not to mention survey independence 
and experiment independence), such a variable redefinition is not 
so useful in seeking a general answer to the question of how many 
dark energy parameters can be characterized.

\section{How many principal components?} \label{sec.eigen} 

Consider the simplest case of a single cosmological parameter 
extended to two parameters.  If the fiducial value of the new parameter is such
that it does not affect the previous parameter sensitivity, then we can treat
this as adding a row and column to the previous Fisher matrix.

The uncertainty on the first variable then grows from $\sigma(v_1)= 
(F_{11})^{-1/2}$ to 
\beq 
\sigma(v_1)=[F_{22}/(F_{11}F_{22}-F_{12}^2)]^{1/2}=(F_{11})^{-1/2} 
\frac{1}{\sqrt{1-r}}, 
\eeq 
where $r=F_{12}^2/(F_{11}F_{22})$ is the covariance between the parameters. 
Covariance blows up parameter errors unless the additional parameter 
is rather uncorrelated with the previous parameter.  This leads to the 
idea of investigating how many uncorrelated parameters can be used to 
characterize the dark energy equation of state. 

The function describing the background behavior of dark energy -- such as 
the equation of state or energy density -- can be decomposed into its 
principal components (PC).  Note that we seek the number of 
components directly informative about the nature of dark energy, so 
while we might obtain, say, 17 PCs to 1\% for distances to $z=0.1-1.7$ 
or 8 PCs of the Hubble parameter, these would require differentiation 
of noisy data to teach us about the dark energy dynamics directly. 
Therefore we concentrate on the dark energy equation of state from 
the start.  

Principal components offer several nice features. 
They are uncorrelated, 
orthogonal, and widely separated by how accurately they can be measured. 
No functional form is assumed, but on the other hand, the eigenmodes 
have no meaning independent of model, survey, or technique (cf.\ 
\S\ref{sec.modelindep}).  
The first principal component gives the quantity most accurately
measured by a particular survey.  Here we extend the work of \cite{hutererpca}
applied to SN measurements by computing the principal components measured
by weak lensing surveys, and also adding the CMB power spectrum 
information.

We assume $w(z)$ to be described by 20 piecewise constant values in redshift
uniformly distributed in redshift between $z=0$ and $z=2$. We form another, 
21st bin
at $2<z<z_{\rm lss}$, but find that its presence has no discernible effect on
the results since dark energy is subdominant at such high redshifts for the 
fiducial $\Lambda$CDM cosmology that we assume.  Once we compute the full
Fisher matrices for SN, WL, and CMB we have an option of adding them.  We then
diagonalize the resulting Fisher matrix for the usual cosmological plus dark
energy parameters, and marginalize over the former. We are left with the 
effective Fisher matrix for the 21 equation of state parameters, which we
diagonalize, and thus obtain the principal components and their
eigenvalues. For more detailed mathematical and practical aspects of this
procedure, see \cite{hutererpca}. 

\begin{figure*}
\psfig{file=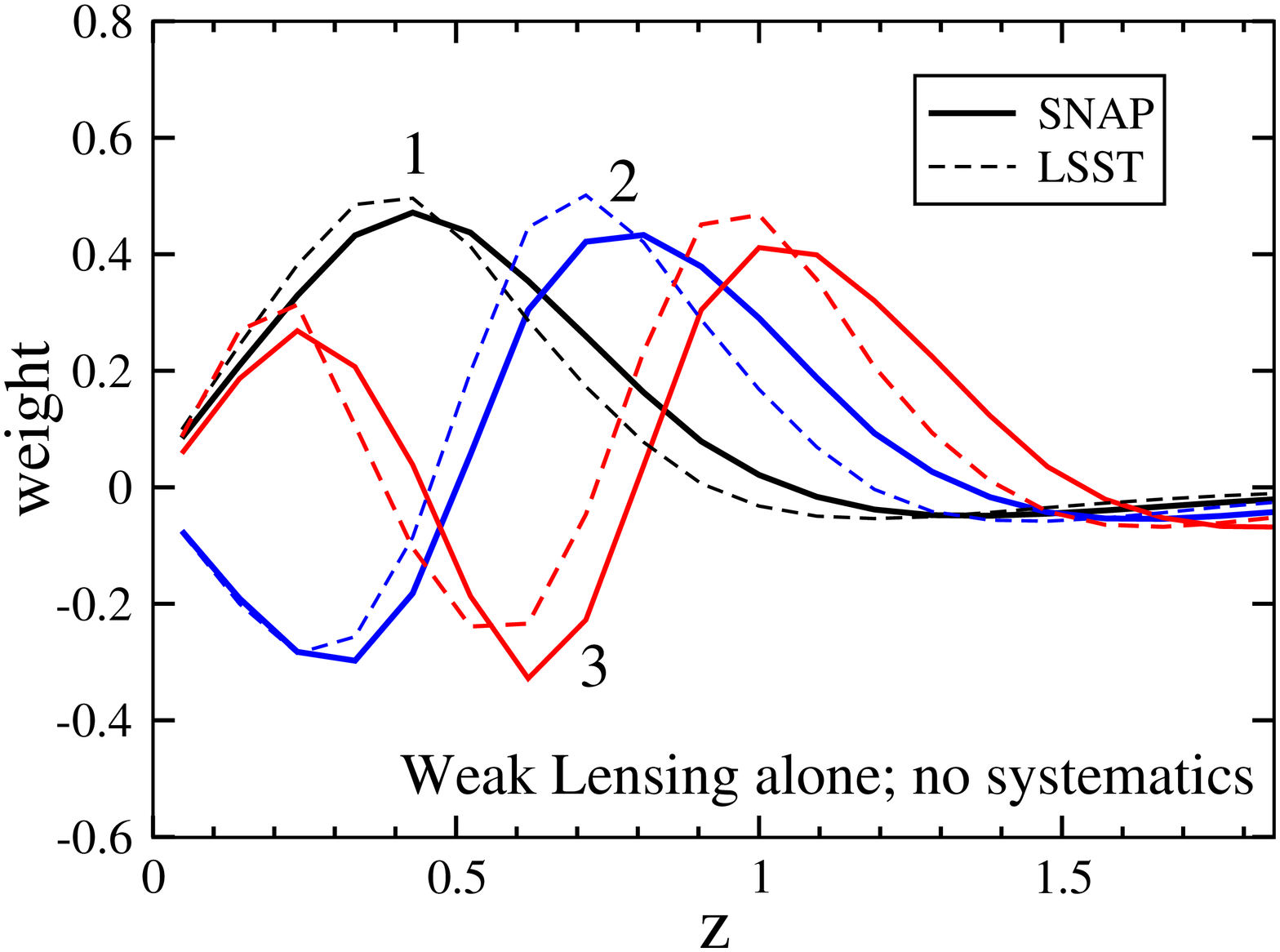, width=2.8in, height=3.5in, angle=-90}\hspace{-0.2cm}
\psfig{file=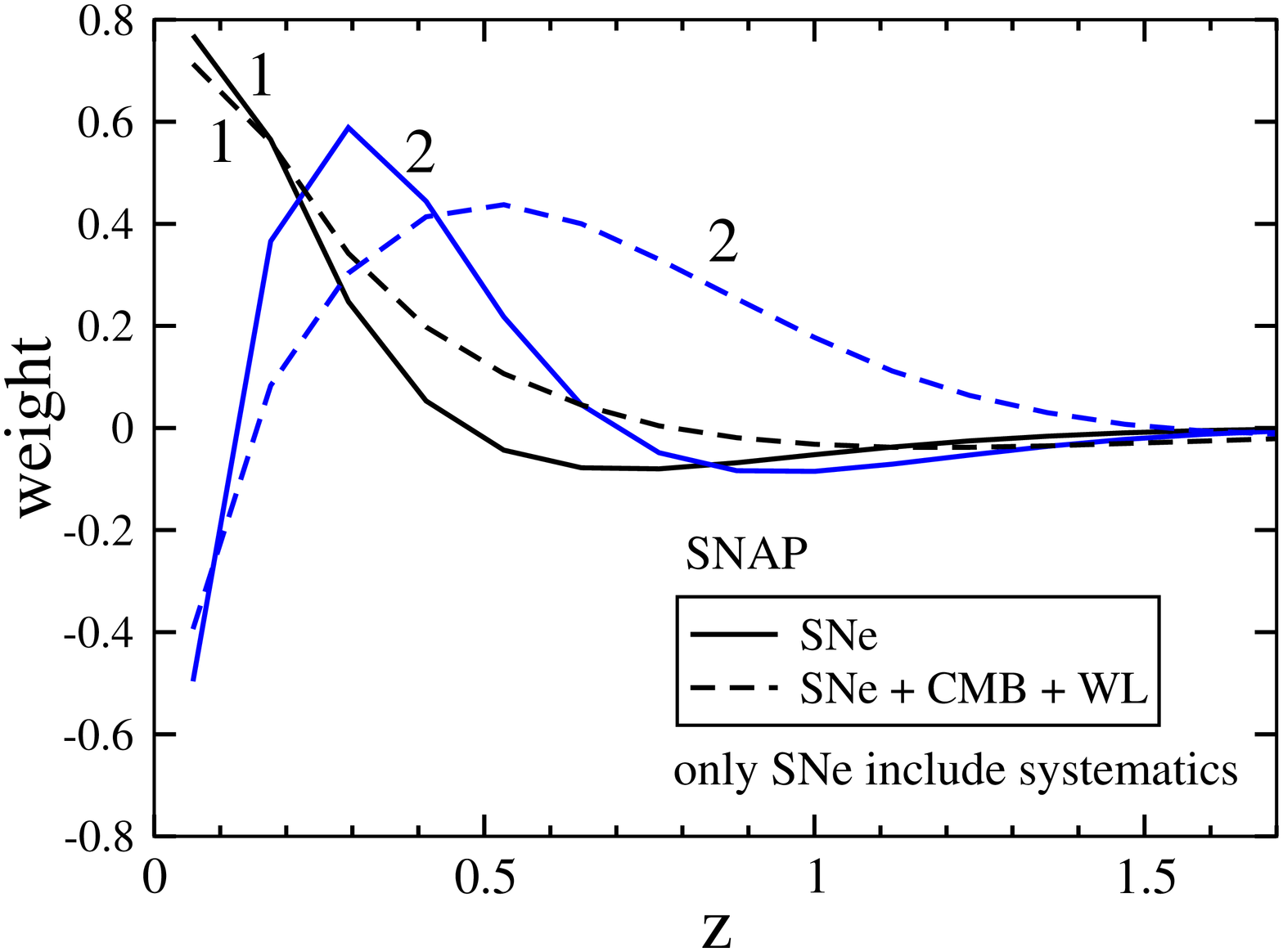,  width=2.8in, height=3.5in, angle=-90}\\[-0.3cm]
\psfig{file=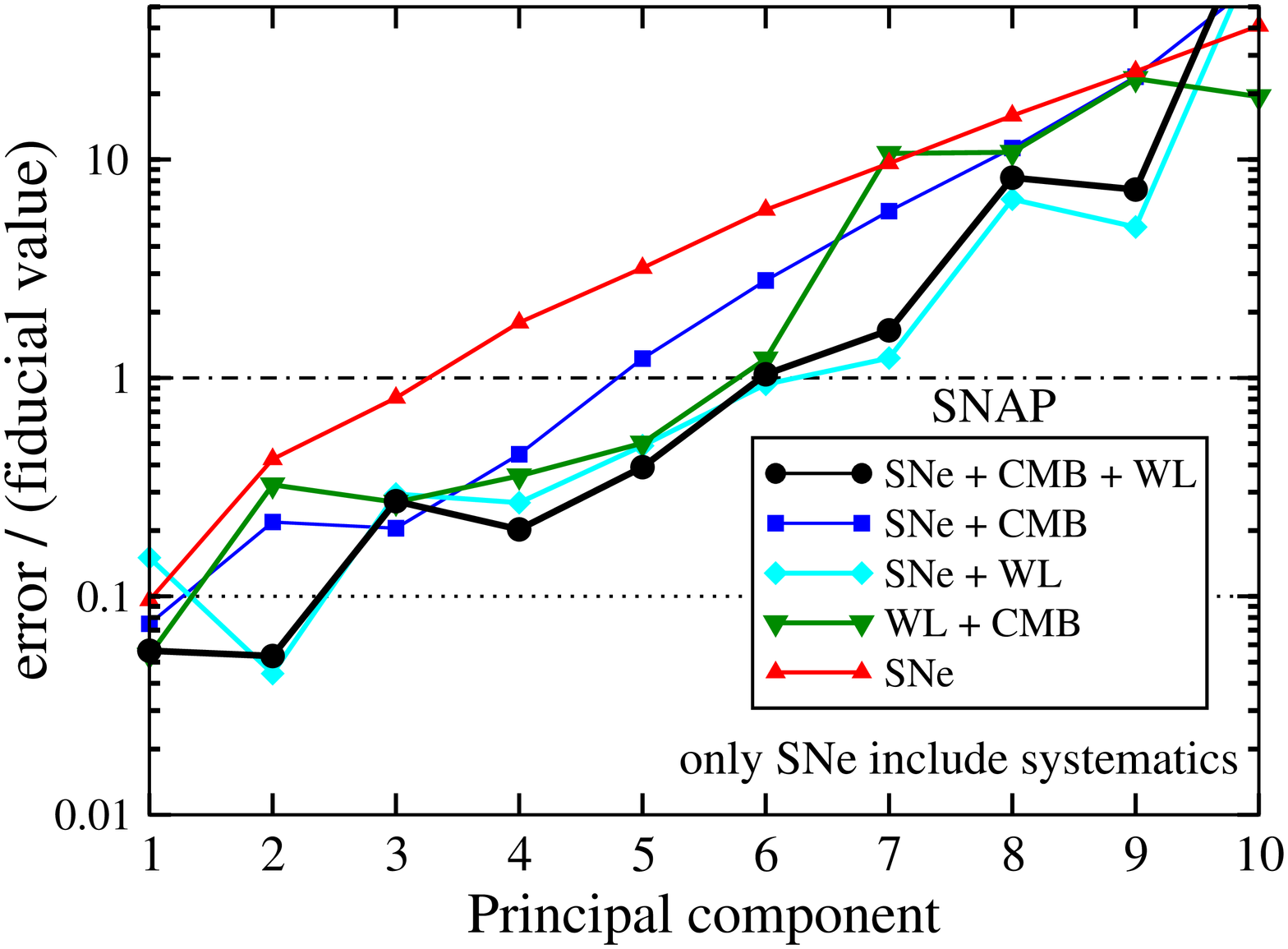, width=2.8in, height=3.5in, angle=-90}\hspace{-0.2cm}
\caption{{\it Top left}: three best measured principal components of $w(z)$ for
the upcoming weak lensing surveys; higher depth and galaxy density 
(SNAP) and wider area (LSST) each contribute well. Note
that the best measured mode peaks at $z\sim 0.4$. {\it Top right}: Two of the
best measured PCs of $w(z)$ expected for a SNAP-type survey for SN alone, and
SN combined with weak lensing and CMB (Planck $d_A(z_{\rm lss})$). The best PC
from SN alone peaks at $z\sim 0$; this becomes $z\sim 0.2$, if $\Omega_M$ is
independently known to better than 0.01 (not shown). Note that the second 
best measured PC is pushed out to higher redshift once the WL and CMB 
information 
is added in. {\it Bottom}: inverse signal-to-noise in measuring the PCs of
$w(z)$; that is, accuracies in measuring the PC coefficients divided by their
fiducial values.  We show the cases of SN alone, SN+CMB, SN+WL, WL+CMB, and the three
probes combined. Throughout, we include the systematics in SN constraints, but not
in WL estimates as the systematics contributions to the latter are highly uncertain.
}
\label{fig:PC}
\end{figure*}

The top left panel in Fig.~\ref{fig:PC} shows the three best measured principal
components of $w(z)$ for weak lensing SNAP and LSST surveys. Note that the best
measured PC peaks at $z\sim 0.4$ -- as expected, the sensitivity of weak
lensing surveys to dark energy is at higher redshifts than that of SN, which
are shown in the top right panel. We also see that while adding the WL+CMB
information does not drastically change the shape of the best SN PC, the
second best-measured PC, and higher ones, are moved to higher redshifts. 

The bottom panel of Fig.~\ref{fig:PC} shows the principal result: the
{\it inverse} signal-to-noise in measurements of each PC for SN  
alone, SN+CMB, SN+WL, WL+CMB, and
the three probes combined.  More precisely, we computed the coefficient of each
principal component entering the fiducial $w(z)$ and its error, and define 
$\sigma_{PC}=$error/coefficient. [Note that it is $\sigma_{\rm
PC}$, and not the raw error, that matters and determines at how many ``sigma''
each PC is detected\footnote{We number PCs in order of
monotonically increasing raw errors, which usually (but not always)
implies monotonically increasing values of $\sigma_{\rm PC}$, depending on the
coefficients entering $w(z)$.}].  To guide the eye we
also plot two horizontal lines corresponding to two criteria for parameter
measurement accuracy: the weak criterion $\sigma_{\rm PC}=1$ (below which we
have a ``1-$\sigma$ detection'' of a given PC) and the strong criterion
$\sigma_{\rm PC}=0.1$ which is precisely equivalent to our 10\% criterion from
Sec.~\ref{sec.success}.

We see that SN measurements alone, with systematics, measure three PCs to
$S/N>1$, and they are further significantly helped by adding the Planck
measurement of $d_A(z_{\rm lss})$ (cf.\ \cite{frieman}).  Weak lensing
power spectra alone, without systematics however, measures four PCs to 
precision satisfying the weak criterion (not shown). Finally, combining
all three probes leads to five (nearly six) PCs measured to $S/N>1$. This
is quite encouraging and shows in a slightly different light that all three
surveys contribute nontrivially to the information content.  However, we 
also find that all three surveys combined lead to only two PCs passing the 
strong criterion, and two more if we dilute it to 25\% precision 
($\sigma_{\rm PC}=0.25$). Our results are in rough agreement with those of Knox
et al.\ \cite{Knox} who also consider the error in measurement of the PCs with
both SN and WL.  (Note that 
taking instead LSST-like WL data does not change the numbers of PCs 
quoted above.) 

These results indicate that it is not just a matter of finding a ``better'' 
parametrization.  Even with allowing $w(a)$ to take on whatever form the 
data best fit, 
next generation data can only generically provide tight constraints on 
two dark energy parameters. 

\section{Conclusion} \label{sec.concl} 

We have examined various characterizations of the dark energy in 
both parametric and non-parametric forms.  Even with observational 
data corresponding to next generation probes of supernova distances, 
weak lensing shear, and the CMB, we find it is extremely challenging 
to give a model independent description of the equation of state 
with more than two parameters: the value at some epoch and a measure 
of its variation.  The good news is that the two parameter 
model in current use is well understood, physically motivated, and 
accurate. 

To some extent this should not be surprising.  If the underlying 
microphysics, e.g.\ potential of the dark energy field, is reasonably 
smooth, described by a few parameters, then we should not expect 
more parameters to be evident in the equation of state.  However we 
note that other microphysics, such as couplings to matter or non-canonical 
sound speed \cite{Hu_GDM,dedeo}, have not been addressed here (although 
if $w\approx-1$ then 
we do not expect the sound speed to have much observable effect
\cite{bean_dore,weller_lewis,cor_speed,hannestad}). 

A more fundamental question to ask is what physics would we learn anyway 
with fitting a third parameter; the physical interpretation of basic 
dynamical characteristics such as an 
equation of state value and its variation is clear, but an extra 
parameter for the sake of having more may not hold much revelation, 
i.e.\ no real qualitative difference.  Furthermore, if we concatenate 
all our cosmological probes together in the quest for one more piece 
of information then we run the risk of losing the ability to test a 
more general framework.  For example, by contrasting the expansion 
history with the growth history \cite{lue_DGP,lingrav,lindeal,knoxtyson}, 
e.g.\ supernovae and weak lensing separately, we can seek whether the 
general concept of an effective equation of state $w(a)$ corresponds 
to the specific origin of a high energy physics scalar field or a 
modification of Einstein gravity.

More elaborate parametrizations have some uses, e.g.\ testing for 
bias in the recovered $w(z)$ (e.g.\ \cite{bassett_compression}) or 
searching for a particular 
property (e.g.\ rapid evolution), but this must be weighed against 
the fact that extra parameters can only be weakly constrained.  In fact, 
even if we reduce a higher parameter model down to two parameters 
by a physically motivated fixing of some parameters, 
in most cases the complementarity of all three cosmological 
probes is essential.  The standard parametrization in terms of 
$w_0$ and $w_a$ is one of the few successful enough to manage a 
generic fit with only a subset of probes; another is the new 
four parameter model introduced, with fixed transition epoch and 
rapidity. 

The principal component decomposition of dark energy equation of
state is a powerful approach that explicitly protects against biases in
parameter determination and separates the accurately measured parameters from
the poorly measured ones. But even the PC method has weaknesses -- the
recovered principal components depend on the survey specifications, and are not
related to fundamental physics of dark energy in any obvious way.  Most 
likely the most powerful approach of characterizing dark energy will be 
testing the data in multiple ways, through both a parametrized function and 
PCA to check for consistency. 

The conclusion is that more than two parameters is not viable for 
a general fit with next generation data if we are interested in
parameters directly informative about the nature and dynamics of
dark energy that do not require further differentiation with
respect to redshift. We can then ask how good 
do we need -- what are the requirements on the data such that we 
would be successful in characterizing 
the dark energy dynamics with a third parameter.  For the power law 
extension, Model~3.1, to estimate $b$ to 0.2 (with a fiducial SUGRA 
model), we require next-next generation improvements in all three 
probes together.  Next-next generation experiments are defined as 
those that determine the distances for $z=0-1.7$ to 0.05\%, 
averaged over redshift bins $\Delta z=0.1$, 
weak lensing shear power spectrum at the level equivalent to that of 
a completely systematics free survey with 10000 square degrees and 
100 galaxies per square arcminute, 
and the CMB distance to last 
scattering to 0.1\%.  For the cubic case, Model~3.2, the requirement 
is next-next generation accuracy on either SN+CMB or WL+CMB. 
For the four parameter dynamics, one can attain successful fits to 
all four parameters with either next-next generation improvements 
to SN or to WL.  However, this was for the most sensitive fiducial 
model and does not hold generally.  For example, next-next accuracy 
on all three probes still fails to fit even three parameters if the 
true model instead possesses, say, $a_t=0.3$ or $\tau=1$. 

So while it is not clear what we would gain in physics insight 
from going beyond two parameters, it is apparent that it would be 
extraordinarily difficult to achieve such characterization.  The 
two basic dynamical properties of the dark energy equation of 
state value and variation seem a reasonable and realistic goal for 
the next generation.

\section*{Acknowledgments} 

This work has been supported in part by the Director, Office of Science,
Department of Energy under grant DE-AC03-76SF00098 and by NSF Astronomy and
Astrophysics Postdoctoral Fellowship under Grant No.\ 0401066.

\end{document}